\documentclass[aps,prl,twocolumn,showpacs,superscriptaddress]{revtex4}
\usepackage{amsmath}
\usepackage{amsfonts}
\usepackage{amssymb}
\usepackage{epsfig}
\usepackage{graphicx}
\newcommand{\be}{\begin{equation}}
\newcommand{\ee}{\end{equation}}
\newcommand{\ba}{\begin{eqnarray}}
\newcommand{\ea}{\end{eqnarray}}
\newcommand{\taul}{\tau_{\rm link}}

\begin{document}

\title{When gel and glass meet: A mechanism for multistep relaxation}

\author{Pinaki Chaudhuri}
\affiliation{Laboratoire PMCN, Universit\'e Lyon 1, 
Universit\'e de Lyon, UMR CNRS 5586, 69622 Villeurbanne, France}

\author{Ludovic Berthier}
\affiliation{Laboratoire des Collo{\"\i}des, Verres
et Nanomat{\'e}riaux, UMR CNRS 5587, Universit{\'e} Montpellier 2,
34095 Montpellier, France}

\author{Pablo I. Hurtado}
\affiliation{Instituto Carlos I de F\'{\i}sica 
Te\'orica y Computacional, 
Universidad de Granada, Granada 18071, Spain}

\author{Walter Kob}
\affiliation{Laboratoire des Collo{\"\i}des, Verres
et Nanomat{\'e}riaux, UMR CNRS 5587, Universit{\'e} Montpellier 2,
34095 Montpellier, France}

\date{\today}

\begin{abstract}
We use computer simulations to study the dynamics of a physical 
gel at high densities where gelation and the glass transition interfere.
We report and provide detailed physical understanding of 
complex relaxation patterns for time correlation functions 
which generically decay in a three-step process.
For certain combinations of parameters 
we find logarithmic decays of the correlators 
and subdiffusive particle motion.
\end{abstract}

\pacs{61.43.Bn, 82.70.Gg, 61.20.Lc}

\maketitle

Various types of phase transformations in disordered materials have been
described in the literature, such as percolation, gelation, or glass
transition~\cite{kob-book}, and it is thus natural to seek novel 
behaviors that emerge from the competition between these different
transformations.
Recently, the structure and dynamics of mixtures of large and
small colloidal particles~\cite{horbach,moreno-mixture}, of materials
with competing lengthscales~\cite{moreno,sperl}, or of systems with
competing interactions~\cite{reviewfuchs,moreno-polymer} has
been studied using a variety of theoretical and experimental techniques.
The motivation to carry out these investigations is of fundamental as well
as  practical nature, since new phenomena can been discovered and disordered
materials with novel properties are created.

Of particular interest are the similarities, differences and competition
between gelation and glass transition which can be observed in a number
of soft materials~\cite{bonn,bartsch,pham,chen}.  Previous
work  focused on particle systems with a hard core repulsion competing
with a very short-range attraction~\cite{reviewfuchs}, which can be
realized experimentally in colloid-polymer mixtures~\cite{pham}. Such
systems can be handled theoretically using liquid state theory for
the structure, and mode-coupling theory for the dynamics, which
then yields detailed predictions for the location and nature of the
dynamic transitions~\cite{reviewfuchs,dawson}.  
Non-ergodic solids are predicted 
both upon compression (glass
physics) or upon increasing the attraction strength (gelation),
with a peculiar dynamical behavior (logarithmic decay of time
correlation functions and sub-diffusive particle displacements) in
the region of parameter space where these transition lines intersect.
A number of numerical simulations~\cite{reviewfuchs,puertas,zac2}
and experiments~\cite{bartsch,pham,chen,mochrie} have confirmed the
overall topology of the phase diagram, with some simulations reporting
the dynamical signatures predicted theoretically~\cite{puertas,zac2}.
However, a detailed physical understanding of these dynamics and an
exploration of their universality for more complex materials has not been
achieved.  Moreover, since mode-coupling singularities are generically
avoided in real materials~\cite{kob-book}, it is important to explore
also the robustness or generality of these results~\cite{dave}, and the
possible deviations or new processes which might emerge in real materials.

We explore these important open directions using computer simulations
of a model system that is a coarse-grained representation~\cite{pablo} of
a transient gel which has been studied in experiments~\cite{porte}. In
this system an equilibrium low-density gel is obtained by adding
telechelic polymers to an oil-in-water microemulsion. Since the polymer
end-groups are hydrophobic, the polymers effectively act as (attractive)
bridges between the oil droplets they connect, whose strength, lengthscale
and typical lifetime can be controlled at will. Denoting by $C_{ij}$ the
number of polymers connecting droplets $i$ and $j$ we have established in
Refs.~\cite{pablo,pablo2} that the following interaction is an effective
coarse-grained representation of this ternary system:

\begin{equation}
V =   \sum_{j > i}
\left( \frac{\sigma_{ij}}{r_{ij}} \right)^{14}  
+ \epsilon_1 \sum_{j > i}
C_{ij} V_{\rm FENE}(r_{ij})
+\epsilon_{0} \sum_{i} C_{ii}.
\label{model} 
\end{equation}

\noindent
The first term is a soft repulsion acting between bare oil droplets,
where $\sigma_{ij} = (\sigma_i + \sigma_j)/2$, $\sigma_i$ is the
diameter of droplet $i$, and $r_{ij}$ is the distance between the
droplet centers.  The second term describes for the entropic attraction
induced by the telechelic polymers, which has the standard ``FENE''
form known from polymer physics~\cite{witten}, $V_{\rm FENE}(r_{ij}) =
\ln ( 1 - (r_{ij}-\sigma_{ij})^2/\ell^2)$, and accounts for the maximal
extension $\ell$ of the polymers. The last term introduces the energy
penalty $\epsilon_0$ for polymers that have both end-groups 
in the same droplet.   
The most drastic approximation of the model
(\ref{model}) is the description of the polymers as effective bonds
between the droplets, which is justified whenever the typical lifetime
of the bonds is much larger than the timescale for polymer dynamics in
the solvent~\cite{porte}.  In order to describe the dynamics of the system,
we use a combination of molecular dynamics to propagate the droplets with
the interaction (\ref{model}), and local Monte Carlo moves with Metropolis
acceptance rates $\tau_{\rm link}^{-1} {\rm min}[1,\exp(\Delta V/k_BT)]$
to update the polymer connectivity matrix 
$C_{ij}$~\cite{pablo,pablo2}. Thus
$\tau_{\rm link}$ is the timescale governing the renewal of the 
polymer network topology.
In order to prevent crystallization we use a polydisperse
emulsion with a flat distribution of particle sizes in the range $\sigma_i
\in [0.75, 1.25]$.

\begin{figure}
\includegraphics[width=85mm,clip]{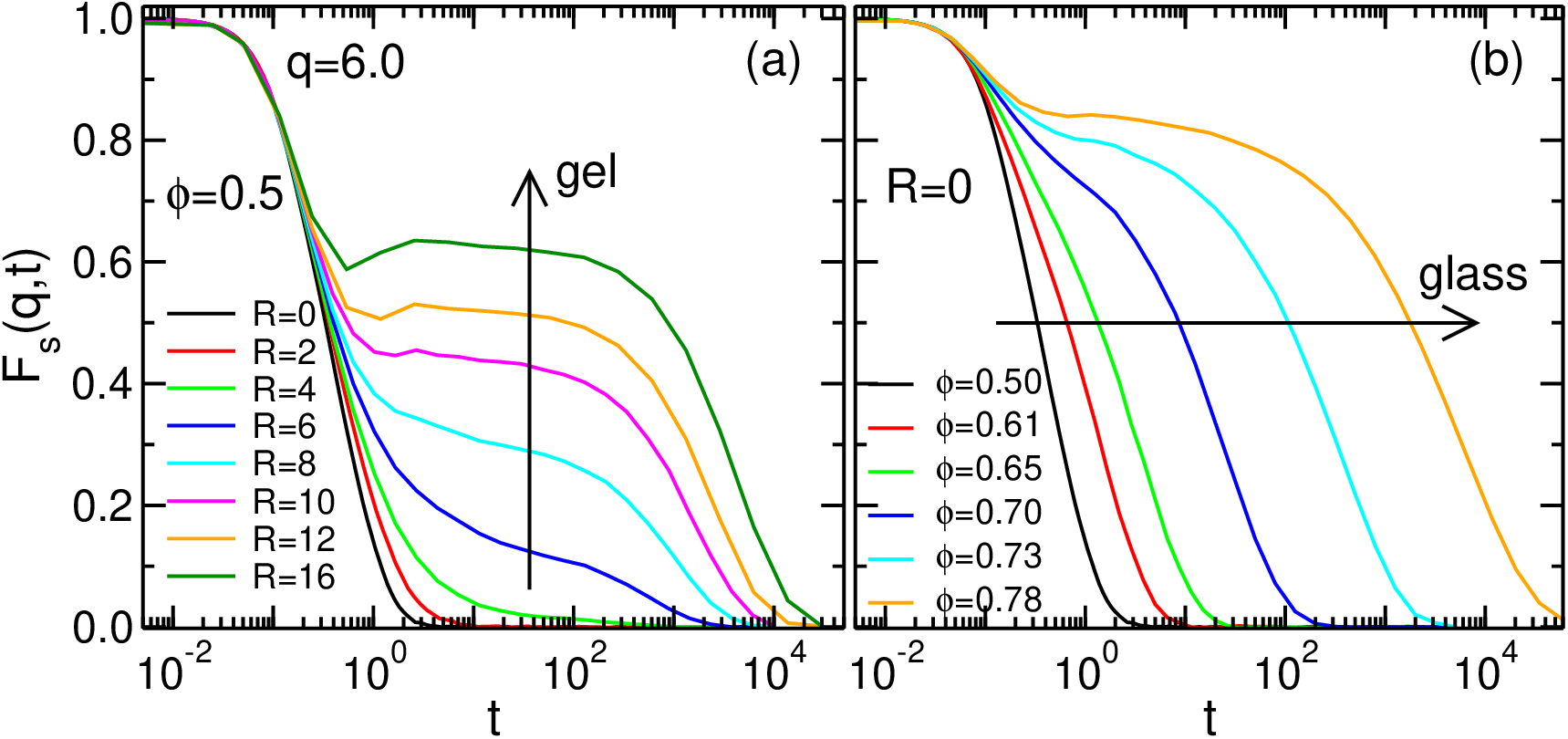}
\caption{\label{glassgel} 
Time dependence of $F_s(q,t)$ for $q=6$. a) Gelation at $\phi=0.5$,
$\tau_{\rm link}=10^2$ and increasing $R$. b) Glass transition at $R=0$
and increasing $\phi$.}
\end{figure}

For moderate volume fraction, $\phi \approx 0.2$, the model behaves as an
equilibrium transient gel with non-trivial dynamics~\cite{pablo} if $R$,
the fraction of polymers per oil droplet, is larger than the percolation
threshold $R_p \approx 2$. In this case, the system is viscoelastic
with a relaxation time set by the bond lifetime $\taul$, and mechanical
strength set by $R$, as illustrated in Fig.~\ref{glassgel}a where we
show the time dependence of the self-intermediate scattering function,

\begin{equation}
F_s(q,t) = \frac{1}{N} \sum_j \exp \left( i {\bf q} \cdot [ {\bf r}_j(t) 
-  {\bf r}_j(0) ] \right),
\label{fsqt}
\end{equation}

\noindent
across percolation for $\phi=0.5$ ($q=6.0$, near the
 main peak in the static structure factor).  A two-step
decay is observed with a plateau height controlled by $R$, and a
slow decay at times $t \approx \taul$. These features
reflect the vibrations of an increasingly stiffer
network of connected particles, followed by a slow reorganization of the
transient network.  In the opposite limit where $R=0$ and the volume
fraction $\phi$ becomes large, the microemulsion becomes a standard
dense glass, Fig.~\ref{glassgel}b.  Here, the two-step decay stems from
particle vibrations within the transient cage formed by the neighbors,
followed by slow structural relaxation.  As usual in this situation,
the relaxation time increases dramatically with $\phi$ while the plateau
height remains constant~\cite{kob-book}.

Our goal in this work is to explore the competition between the
two well-documented phenomena illustrated in Fig.~\ref{glassgel}.
The space of control parameters is large, so we fix~\cite{pablo} 
$\{\ell = 3.5, k_B T = \epsilon_0 = 1, 
\epsilon_1 = 50\}$ and vary $\{ \phi, R, \tau_{\rm
link}\}$.  We successively describe the effect of (i) increasing
the density of an equilibrium gel, (ii) adding attractive interactions to
a viscous close to the glass transition, (iii) changing the bond
lifetime of a system close to dynamic arrest to estimate the relative
importance of bonding and steric hindrance for the
relaxation.

\begin{figure}
\psfig{file=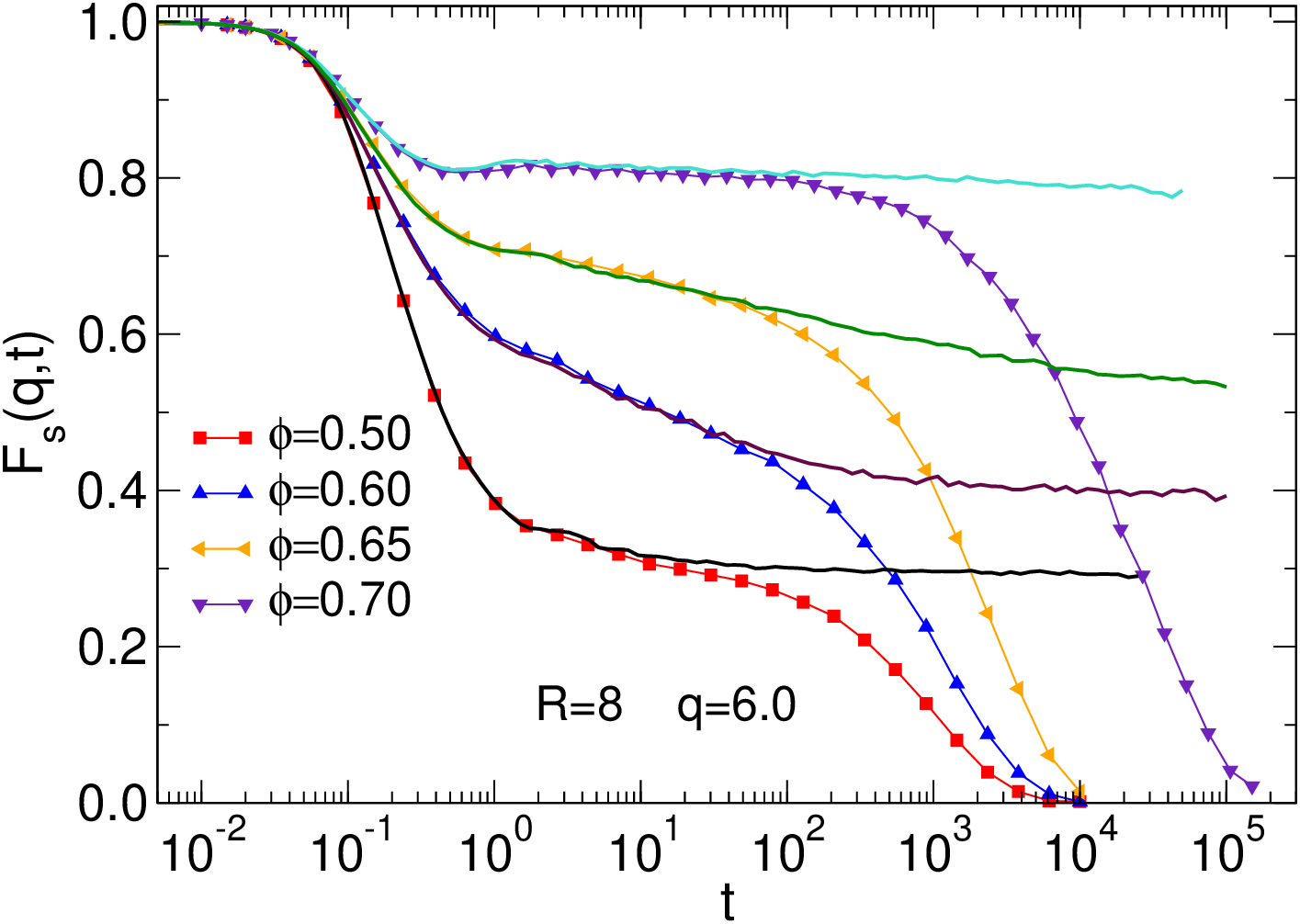,width=8cm,clip}
\caption{\label{results} 
Compressing an equilibrium transient gel with $R=8$ towards the glass
transition.  The symbols are for $\tau_{\rm link}=10^2$ and the full
lines are for frozen bonds, $\taul = \infty$.  The dynamical evolution
is a non-trivial mixture of both gelation and glass transition with a
logarithmic decay towards a plateau at intermediate volume fraction.}
\vspace*{-5mm}
\end{figure}

We start our exploration in Fig.~\ref{results} which shows the evolution
of the equilibrium gel dynamics, $R=8$, $\tau_{\rm link}=10^2$, as it
is compressed from $\phi=0.5$ towards the glass transition. Contrary to
the extreme cases shown in Fig.~\ref{glassgel}, here both the relaxation
time and the plateau height increase simultaneously. Thus these results
are not explained by gelation or the glass transition alone, but truly
result from the non-trivial effect of their competition.  This is directly
demonstrated by increasing, for the same state points, the bond lifetime
$\taul$ to a very large value.  For $\phi=0.5$, the correlator quickly
decays to the plateau and then becomes completely arrested, showing
that at low density only the transient network physics is at play. At
larger density, $\phi=0.60$ and 0.65, a very slow decay towards a plateau
is observed.  In particular, for $\phi=0.65$ the data follow a nearly
logarithmic decay over about 5 decades in time, a behavior which is seen
neither for the gel nor the glass alone. This means that the 
elastic relaxation of
the network is considerably slowed down by crowding effects.
Interestingly, qualitatively similar experimental observations have
recently been reported in  micelles~\cite{chen} and attractive
nanoparticles~\cite{mochrie}.  At even larger volume fractions,
$\phi=0.7$, $F_s(q,t)$ decays again in a two-step process,
but the slow decay is now controlled both by the bond-lifetime (as in
gels) and by density (as glassy liquids).  Although the
dynamics at large $\phi$ and $R$ is a two-step process similar to the
one of the glassy fluid at $R=0$, the effect of $\taul$  on the second
decay establishes its very different nature. 
While crowding alone is
responsible for the slow dynamics near $R=0$, bond-lifetime and network
reorganization control the dynamics at large $R$.
This physical distinction is
reminiscent of the ``bonded glass'' and ``repulsive glass'' nomenclature
introduced for attractive colloids~\cite{poon}.
It is between these two regimes that 
multistep relaxations can be observed. 

\begin{figure}
\includegraphics[width=85mm,clip]{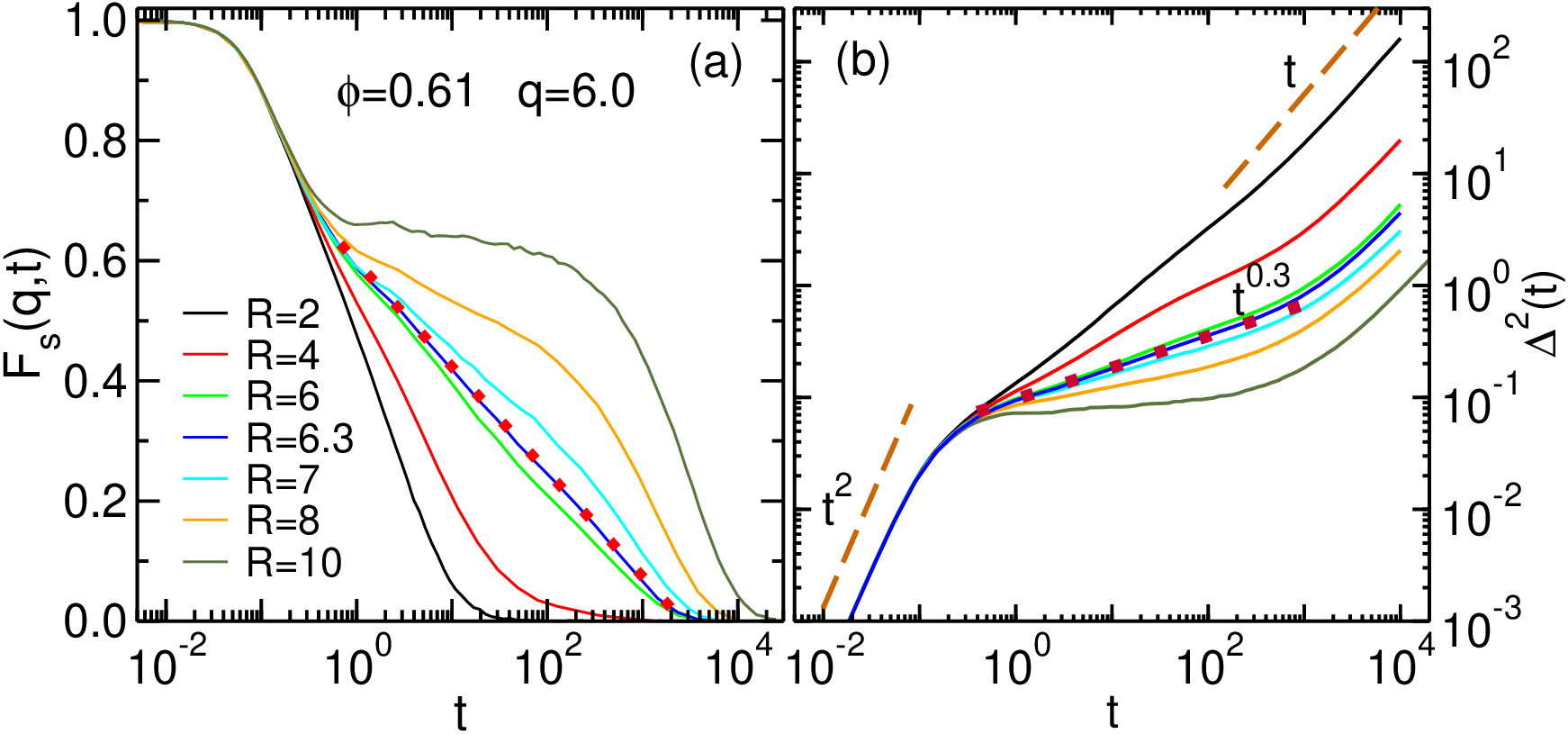}
\caption{\label{fslog}
Effect of increasing the attractive interactions, $R$, in a dense
repulsive fluid, $\phi=0.61$, for a finite bond lifetime, $\taul=10^2$
on (a) $F_s(q,t)$ and (b) $\Delta^2(t)$.  For intermediate $R$
values, a nearly logarithmic time decay of $F_s(q,t)$ is observed
(symbols), associated with a sub-diffusive behavior of single particle
displacements.}
\end{figure}

Alternatively, gel and glass can meet by increasing
the attraction in dense systems, as usually done in polymer-colloid
mixtures~\cite{pham,puertas}. Note that in our model we can increase
the attraction between droplets, $R$, without simultaneously
increasing the bond lifetime $\taul$, while these quantities are 
tightly coupled in
attractive colloids~\cite{pham,zac2,puertas}.  This allows us to
disentangle static from dynamic effects.  In Fig.~\ref{fslog}
we show the effect of increasing $R$ in a dense fluid at $\phi=0.61$
using a constant bond lifetime $\tau_{\rm link}=10^2$.  We show both the
evolution of $F_s(q,t)$, and of the mean-squared displacement,

\begin{equation}
\Delta^2(t) = \frac{1}{N} \sum_i \left\langle 
| {\bf r}_i(t)  - {\bf r}_i(0) |^2 \right\rangle.
\end{equation} 

As for the low density system in Fig.~\ref{glassgel}, increasing the
attractive interaction allows the system to cross the percolation
line and to become viscoelastic, i.e. $F_s(q,t)$ decays in two steps.
However, the proximity of the glass transition makes the dynamic
evolution more complex, as this produces, at intermediate values of $R
\approx 5-7$ a very slow decay of the time correlation function, which
can be empirically described by a logarithmic decay, Fig.~\ref{fslog}a.
This behavior resembles the ones reported in numerical work for colloids
with short-range attraction~\cite{zac2,puertas}.  The mean-squared
displacement also evidences deviations from a simple two-step process,
as particle displacements appear to be transiently sub-diffusive at
those intermediate $R$ where logarithmic decay is observed.  Again this
behavior can not be accounted for by a simple ``superposition'' of gel
and glass dynamics.

\begin{figure}
\includegraphics[width=80mm,clip]{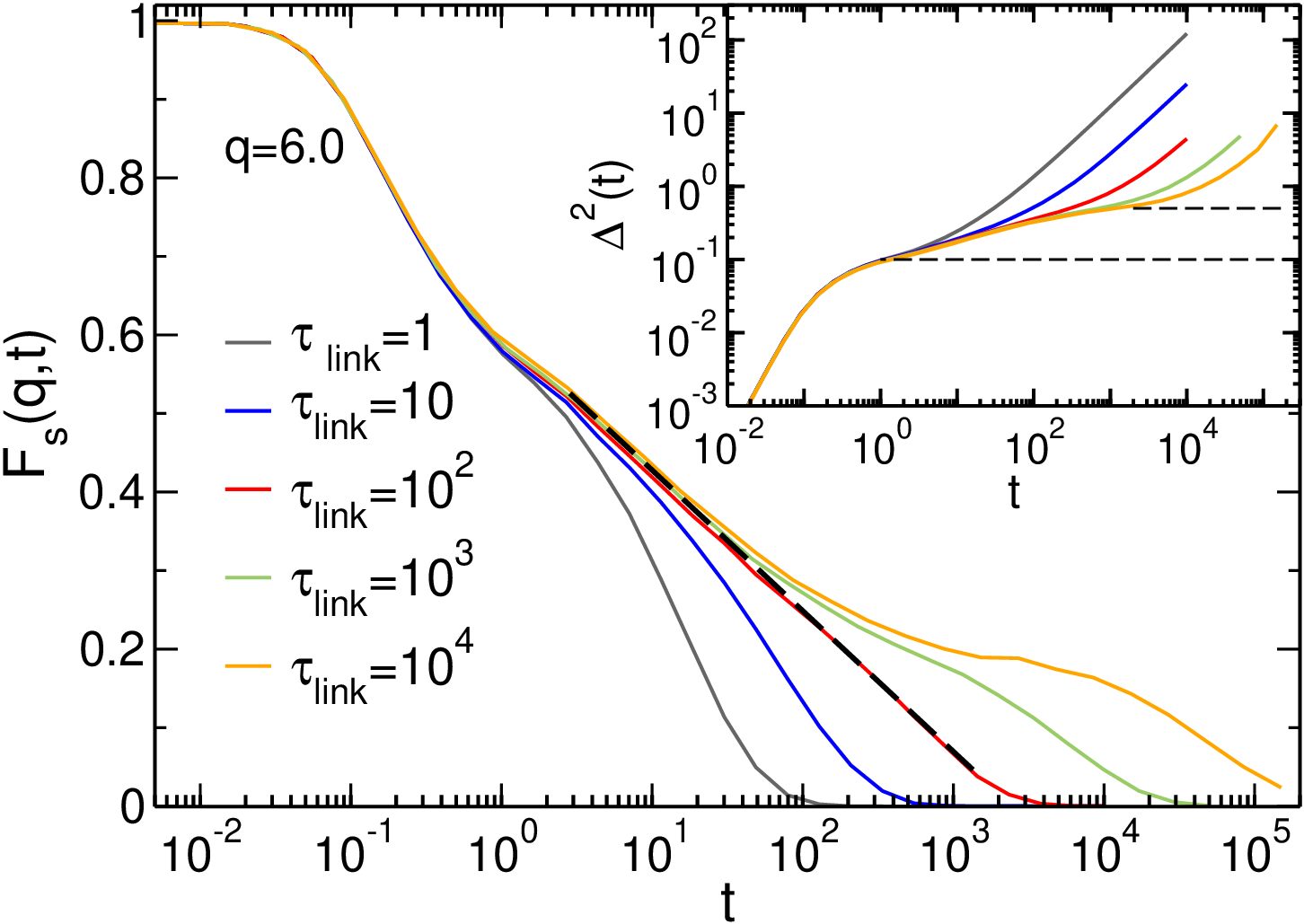}
\caption{\label{deconstruct}
Increasing the bond lifetime $\taul$ from 1 to $10^4$ at $\phi=0.61$ and
$R=6.3$ produces a crossover from nearly logarithmic 
(dashed line)
to a three-step
process when timescales are well-separated. The inset shows the
mean-squared displacement for the same parameters, revealing the existence
of the distinct lengthscales (dashed lines) controlling the dynamics.}
\vspace*{-5mm}
\end{figure}

We explore further the dynamics at those state points where the
dynamics is non-trivial, and make use of the flexibility offered by the
present model to change attraction and bond-lifetime independently.
In Fig.~\ref{deconstruct}, we show the effect of increasing $\taul$
on $F_s(q,t)$ and $\Delta^2(t)$ for constant values of $R=6.3$ and
$\phi=0.61$, i.e. where the logarithmic decay was most prominent in
Fig.~\ref{fslog}.  We now realize that the logarithmic decay seen
for $\tau_{\rm link}=10^2$ is in fact a very specific instance of
a more generic three-step decay of time correlation functions which
reflects the double localization of the particles within their cages
and within the transient particle network.  This three-step process
is also clear from  the behavior of the mean-squared displacements
(inset of Fig.~\ref{deconstruct}), which reveals the existence of two
lengthscales controlling the dynamics of this system. A first plateau is
observed for $\Delta^2 \approx 0.1$, which corresponds to the typical
cage size in repulsive glasses~\cite{kob-book}, and a second plateau
is observed near $\Delta^2 \approx 0.5$, which corresponds to the
localization of the particle due to the presence of the percolating
polymer network~\cite{pablo}. The logarithmic decay is only observed if
a non-generic combination of timescales and lengthscales combines both
slow processes in a single, nearly logarithmic one. In fact our data
suggest that the {\it generic} situation should be the occurrence of a
three-step process corresponding to well-separated bond/cage relaxation.
To the best of our knowledge, there is no experimental report of such
a three-step decay of time correlation function.  We suggest that
compressing the transient gel system of Ref.~\cite{porte} is a possible
route to such observations.

\begin{figure}
\includegraphics[width=80mm]{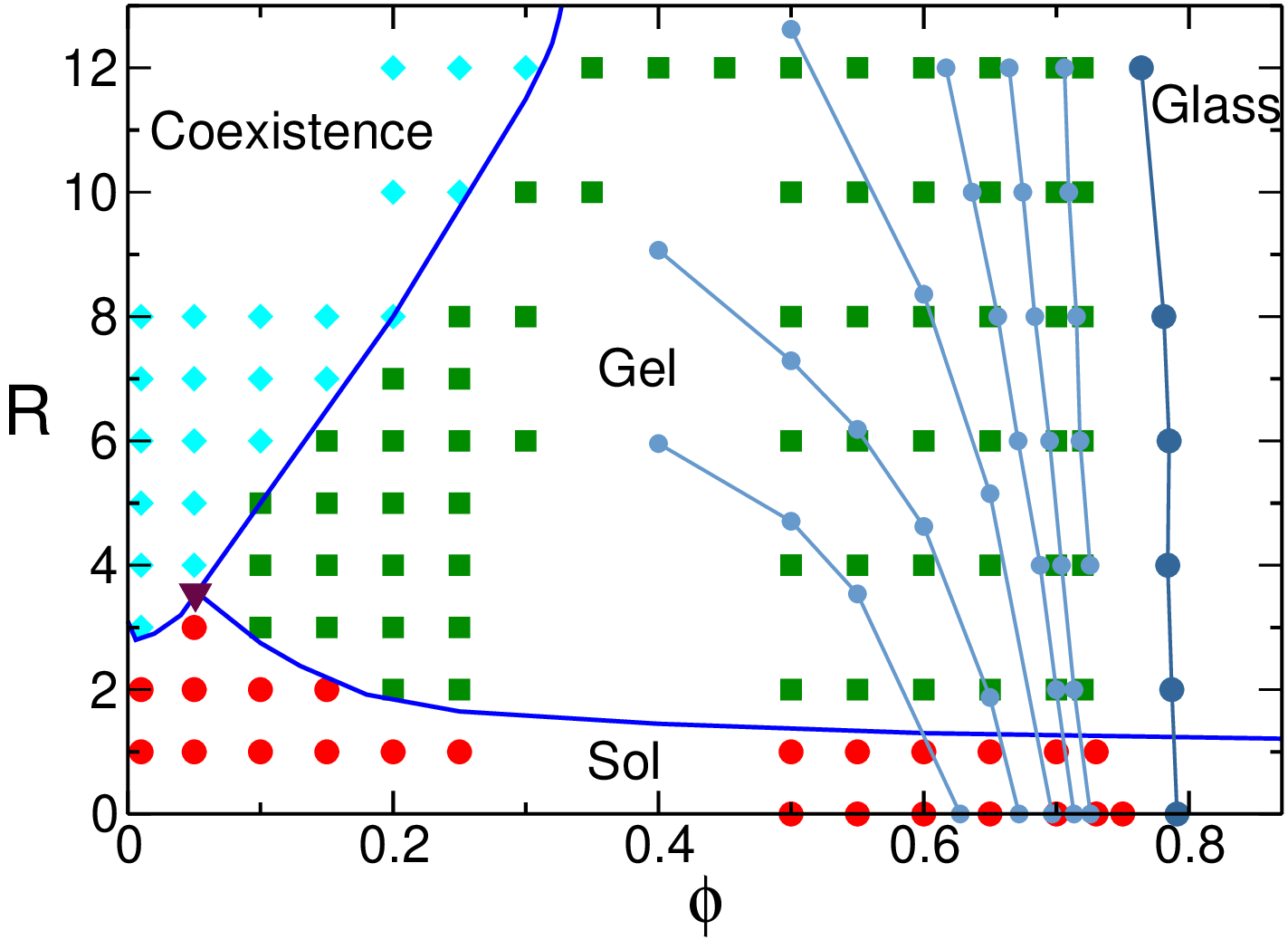}
\caption{\label{phase}
Phase diagram for $\tau_{\rm link}=1.0$:
Coexistence region ($\diamond$), sol
($\bigcirc$) and gel ($\square$) phases, and measured glass
line ($\bigtriangleup$), determined from 
a Vogel-Fulcher fit of the relaxation time.
Also shown are the iso-$\tau$
lines ($\bigtriangledown$) for $\tau=10,30, 10^2, \cdots ,10^4$.} 
\end{figure}

We now summarize our findings in a phase diagram in Fig.~\ref{phase}.
The low density part, $\phi < 0.2$ was described earlier~\cite{pablo}.
It contains a sol phase if $\phi$ and $R$ are small, and a phase separated
region when attraction is increased at low $\phi$. There is a
large region (currently explored in experiments~\cite{porte,mora}) where a
homogeneous, equilibrium, transient gel is formed. In the present work we
have explored the interplay  between  the glass transition at large volume
fraction and gelation at large $R$, where we discovered new dynamical
phenomena.  The iso-relaxation time lines reported in Fig.~\ref{phase},
obtained for $\taul =1.0$ and defined by $F_s(q,\tau)=0.03$, show
that the dynamics slows down both by increasing $\phi$ or $R$.  In the
region where both $R$ and $\phi$ act as a source for slow dynamics we
generically obtain a three-step relaxation process, which produces for
specific combinations of $R$, $\phi$, and $\taul$, 
a logarithmic decay towards a
plateau followed by a slower decay, as in Fig.~\ref{results}, or a fully
logarithmic decay as in Fig.~\ref{fslog}.  The present model thus captures
a broad range of behaviors, and also predicts new types of relaxations.

The phase diagram in Fig.~\ref{phase} shows that, in the gel phase, ergodic
behavior is found even if the attraction gets very large. For $\phi=0.5$,
we have done simulations up to $R=50$ and still found ergodic behavior,
the relaxation time increasing smoothly with $R$. 
Thus we find no evidence of an `ideal' gel phase, as predicted
theoretically for attractive particles~\cite{dawson,reviewfuchs,fuchs},
although of course the relaxation time can get very large if both $R$ and
$\taul$ increase.  This suggests a fundamental difference between gelation
and glass transition because in (fragile) glass-forming materials, the
predicted mode-coupling singularity is also avoided, but it is believed to
be replaced by a truly divergent timescale~\cite{kob-book}.  As suggested
also from simulations of particles with attractive patches~\cite{patchy},
gels could be the analog of Arrhenius (strong) glasses with no finite
temperature singularity.

One important finding of our study is the prediction of a generic
three-step decay of time correlation functions, which has not yet been
directly observed experimentally. As shown in Fig.~\ref{deconstruct}, this
requires the existence of two distinct lengthscales and well-separated
timescales for cage/bond relaxation. 
We suggest that the observation of a two-step process
for yielding in a sheared colloid-polymer mixture~\cite{yield} might
be a rheological analog of Fig.~\ref{deconstruct}, which certainly
calls for further investigations.  
Even more complex relaxation dynamics could potentially emerge 
in transient gels made with polymer mixtures having distinct 
lengthscales, thus opening the door to the creation of new materials.
 
\acknowledgments
Financial support from Spanish project FIS2009-08451 and ANR's TSANET
and SYSCOM is acknowledged.

\end{document}